\documentclass[twocolumn,twoside]{article}

\def\R{{I\!\! R}}

\newcommand{\dam}{\partial_\nu\partial^\nu}
\newtheorem{theorem}{Theorem}
\begin{document}
\markboth{H. Gottschalk}{Stability of one particle states}
\thispagestyle{empty}
\title{ On the stability of one particle states generated by quantum
fields fulfilling Yang-Feldman equations}
\author{
 Hanno Gottschalk,\\
Institut f\"ur angewandte Mathematik, Universit\"at Bonn\\
Wegelerstr. 6, D-53115 Bonn, Germany\\ e-mail:
gottscha@wiener.iam.uni-bonn.de} \maketitle
\begin{abstract}
{\small We prove that for a Wightman quantum field $\phi(x)$ the
assumptions (i) positivity of the metric on the state space ${\cal
H}$ of the theory (ii) the asymptotic condition in the form of
Yang-Feldman equations and (iii) Klein-Gordon equation for the
outgoing field imply that the states generated by application of
the asymptotic fields to the vacuum are stable. We prove by a
counter example that this statement is wrong in the case of
quantum fields with indefinite metric. }
\end{abstract}

\noindent {\small {\bf Mathematical subject classification 2000:}
81T05 \\ \noindent {\bf Key words:} {\em Quantum field theory,
Yang-Feldman equations, mass shell singularities.} }
\section{Introduction}
It is well-known that in positive metric quantum field theory
(QFT) the decay reaction of a particle with mass $m>2\mu$ into two
particles of mass $\mu$
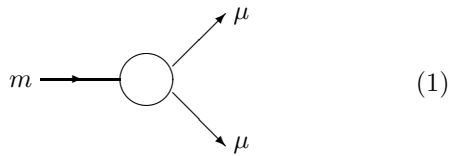
\begin{equation}
\label{1eqa}
\begin{array}{c}
\begin{picture}(100,50)(0,0)
\put (50,25){\circle{20}} \put(10,25){\line(1,0){30}}
\put(25,25){\vector(1,0){1}} \put(60,30){\vector(1,1){20}}
\put(60,20){\vector(1,-1){20}} \put(-2,22){$m$} \put(83,48){$\mu$}
\put(83,-1){$\mu$}
\end{picture}
\end{array}
\end{equation}
is prohibited kinematically, since the one particle in- state has
sharp mass $m$ and the two particle out- state has continuous mass
spectrum. In this note we prove a generalization of this statement
which shows that in positive metric quantum field theory the
asymptotic condition in form of the Yang-Feldman
equations\cite{YF} and the Klein-Gordon equations for the
asymptotic fields are in contradiction with a nontrivial
transition of a one particle in-state into an arbitrary state (not
necessarily a multi-particle state). No restrictions on the
spectrum of this state have to be assumed. This might be of
particular interest in QFTs containing several species of
particles with the same mass or QFTs which are not asymptotically
complete \cite{Ri}.

Positivity crucially enters into the proof of this statement and
we show by a counter example that there exist quantum fields with
indefinite metric such that even a non-trivial decay reaction
(\ref{1eqa}) is possible.

The article is organized as follows: In the next section we state
our assumptions and we give a suitable formulation of Yang-Feldman
equations. In section 3 we state and prove the main theorem.
Section 4 is devoted to the counter example for the case of QFTs
with indefinite metric.

\section{Preparations}
Let $\phi(x)$ be a relativistic Wightman field (cf. \cite{SW})
acting on the Hilbert space ${\cal H}$ with vacuum $\Psi_0$. The
theory is allowed to contain an arbitrary number of quantum
fields, in particular we do not assume cyclicity of the vacuum
w.r.t. $\phi(x)$. For simplicity we assume $\phi(x)$ to be
Hermitean and scalar (these assumptions however are not crucial).
$j(x)$ we denote the current fields of $\phi(x)$, i.e.
\begin{equation}
j(x)=(-\dam-m^2)\phi(x).
\end{equation}

Next we come to the question, how to define asymptotic states for
the field $\phi(x)$ without assuming an {\em isolated} mass-shell
of mass $m$ in the spectrum of the energy-momentum operator ${\sf
P}$ (this automatically would imply stability of one particle
states). Haag-Ruelle \cite{Ha,He,Ru} theory can not be applied in
this situation. We help ourselves by {\em postulating} the
asymptotic condition \cite{LSZ}in the following rather weak form
based on the Yang-Feldman equations: Let $f\to\phi^{\rm
in/out}(f)\Psi_0$ be a vector valued distribution which assigns to
a test function $f$ in the space of rapidly decreasing test
functions ${\cal S}={\cal S}(\R^4)$ a vector in the state space
${\cal H}$ of the theory. We postulate that the Yang-Feldman
equations for $\phi$ hold for certain matrix elements, in
particular
\begin{eqnarray}
\label{6eqa} &&\left\langle
\phi(x)\Psi_0,\Psi\right\rangle=\nonumber\\ &&\Big\langle
\left[(-\dam-m^2)^{-1}_{\rm ret/adv}*j(x)+\phi^{\rm
in/out}(x)\right]\nonumber\\ &&\times\Psi_0,\Psi\Big\rangle
~~\forall \Psi\in{\cal H}
\end{eqnarray}
(here it would be sufficient to take $\Psi$ from a dense subset of
${\cal H}$). In order to give a precise sense to Eq. (\ref{6eqa})
we have to assume that the distributional convolution of $j(x)$
with the retarded/advanced Greens function $(-\dam-m^2)^{-1}_{\rm
ret/adv}$ is well-defined in (\ref{6eqa}). A suitable technical
formulation is the following: Let $\hat w(k)=\hat w(k,\Psi)$ be
defined as
\begin{equation}
\label{7eqa} \hat w(k)=\int_{\R^4}e^{ix\cdot k}\left\langle
j(x)\Psi_0,\Psi\right\rangle dx
\end{equation}
We have to make sure that the product of $\hat w(k)$ and the
Fourier transform $1/(k^2-m^2\mp ik^0\epsilon)$ of
$(-\dam-m^2)^{-1}_{\rm ret/adv}$ is well defined. A sufficient and
essentially necessary condition for this is that $\hat w(k)$ is
continuously differentiable in the variable $\kappa=k^0-\omega$ in
an open neighborhood of $\kappa=0$, $\omega=\sqrt{|{\bf
k}|^2+m^2}$. We shall assume this in the following. Furthermore,
we assume the Klein-Gordon equation for the outgoing field applied
to the vacuum
\begin{equation}
\label{8eqa} (-\dam-m^2)\phi^{\rm out}(x)\Psi_0=0.
\end{equation}
Using the Yang-Feldman equations it is not difficult to see that
this implies the Klein-Gordon equation also for the incoming field
applied to the vacuum.
\section{The theorem}
We now can formulate the main result of this note:
\begin{theorem}
 In any quantum field theory with positive metric
the assumptions of Section 2 concerning the Yang-Feldman equations
(\ref{6eqa}) and the Klein-Gordon equation for the outgoing field
(\ref{8eqa}) imply stability of the one particle states of the
incoming and outgoing fields in the sense
\begin{equation}
\label{9eqa}
 \phi^{\rm in}(x)\Psi_0=\phi^{\rm out}(x)\Psi_0.
\end{equation}
\end{theorem}
The rest of this section is devoted to the proof of this theorem.
We suppose that Eq. (\ref{9eqa}) does not hold to derive a
contradiction. If Eq. (\ref{9eqa}) does not hold then $\exists
\Psi\in{\cal H}$ such that
\begin{equation}
\label{10eqa} \langle\phi^{\rm
in}(x)\Psi_0,\Psi\rangle\not=\langle\phi^{\rm
out}(x)\Psi_0,\Psi\rangle.
\end{equation}
We set
\begin{eqnarray}
\label{12eqa} \hat W^{\rm  -/in/out}(k)&=& \int_{\R^4}e^{ix\cdot
k}\nonumber\\ &\times&\Big\langle \phi^{\rm -/in/out}(x)\Psi_0,
\Psi\Big\rangle ~dx,\nonumber\\
\end{eqnarray}
where $-$ stands for the local field. We  then get for the
Yang-Feldman equations (\ref{6eqa})
\begin{equation}
\label{13eqa} \hat W(k)= {\hat w(k)\over ( k^2-m^2)\mp
ik^0\epsilon}+\hat W^{\rm in/out}(k)
\end{equation}
and further more $\hat W^{\rm in}(k)\not=\hat W^{\rm out}(k)$, cf.
Eq. (\ref{10eqa}).

{\bf Step 1)} We want to show that by (\ref{10eqa})  the matrix
element for the current $j$ can not vanish identically 'on shell':
Thus suppose that $\hat w(k)=0$ whenever $\kappa=0$. Since $\hat
w(k)$ is continuously differentiable in an neighborhood of
$\kappa=0$ we get that $\hat w(k)=\kappa b(k)$ for some function
$b(k)$ which is continuous on a neighborhood of $\kappa=0$. Using
that $\kappa b(k)/(\kappa+ik^0\epsilon)=\kappa
b(k)/(\kappa-ik^0\epsilon)=b(k)$ holds in the sense of tempered
distributions, we get
\begin{equation}
\label{14eqa} {\hat w(k)\over ( k^2-m^2)- ik^0\epsilon}={\hat
w(k)\over ( k^2-m^2)+ ik^0\epsilon}
\end{equation}
and hence, by Eq. (\ref{13eqa}), $\hat W^{\rm in}(k)=\hat W^{\rm
out}(k)$ in contradiction with Eq.(\ref{10eqa}). Here we also used
that $\hat w(k)$ vanishes on the backward mass-shell due to the
assumptions on the spectrum of ${\sf P}$ (\cite{SW}). Thus, there
must be a Schwartz test function $g:\R^3\to \R$ s.t.
\begin{equation}
\label{15eqa} \tilde w(\kappa)=\int_{\R^3}{\hat
w(\kappa+\omega,{\bf k})\over \kappa+2\omega} g({\bf k}) ~d{\bf k}
\end{equation}
is continuous differentiable and not equal to zero in a
neighborhood of $\kappa=0$.

{\bf Step 2)} Now let $\hat \varphi:\R\to\R$ be an antisymmetric
function ($\hat\varphi(\kappa)=-\hat \varphi(-\kappa)$) s.t.
$\hat\varphi(\kappa)=\kappa$  for $\kappa \in[-1,1]$,
$2\geq\hat\varphi(\kappa)\geq0$ for $\kappa\geq0$ with support in
$[-2,2]$. We set
$\hat\varphi_n(\kappa)=\hat\varphi(n\kappa)/\tilde w(\kappa)$
which is a continuously differentiable function for sufficiently
large $n$. We observe that for such $n$
\begin{eqnarray}
\label{16eqa} &&\left|\int_{\R^4}\hat W(k)\hat
\varphi_n(k^0-\omega)g({\bf k})~dk\right|\nonumber\\
&=&\left|\int_{\R}\hat\varphi_n(\kappa){\tilde w(\kappa)\over
\kappa+i\epsilon}~d\kappa\right|\nonumber\\ &=&\left|\int_{\R}
{\hat\varphi(n\kappa)\over\kappa}d\kappa\right|\nonumber\\
&\geq&\int_{-1/n}^{1/n} n ~d\kappa=2
\end{eqnarray}
In the first  step we have used the Yang-Feldman equation in the
form (\ref{13eqa}) for the out-case taking into account $\hat
\varphi_n(\kappa)\hat W^{\rm out}(k)=0$ by Eq. (\ref{8eqa}) and
the spectral condition \cite{SW}. In the second step we used that
$\hat\varphi_n(\kappa)$ vanishes for $\kappa=0$ and we used the
definition of $\hat\varphi_n(\kappa)$. In the third step we used
the fact that by definition of $\hat\varphi$ we get
$\hat\varphi(n\kappa)/\kappa\geq 0$ for all $\kappa\in\R$ and
$\hat\varphi(n\kappa)/\kappa=n$ for $\kappa\in[-1/n,1/n]$. The
last step is trivial.

{\bf Step 3)} We define the following sequence of test functions
$f_n$:
\begin{equation}
\label{17eqa} f_n(x)=(2\pi)^{-4}\int_{\R^4}e^{-ik\cdot
x}\hat\varphi_n(k^0-\omega)g({\bf k})~dk.
\end{equation}
 The $f_n$ might not be Schwartz test functions (this happens if $\tilde w(\kappa)$ is not infinitely often differentiable in a neighborhood of zero). But evaluation of all distributions we consider will be well-defined for the test functions $f_n$ and the argument we give below can be made fully rigorous by smearing out the Fourier transform of the $f_n$. We leave the details to the reader.
We get by construction
\begin{equation}
\label{18eqa} 2\leq \lim_{n\to\infty}\left|\left\langle
\phi(f_n)\Psi_0,\Psi\right\rangle\right|.
\end{equation}
The positivity of a  Hermitean inner product
$\langle\cdot,\cdot\rangle$ is equivalent with the Cauchy-Schwarz
inequality for this product. Applying this inequality to the above
equation we get
\begin{eqnarray}
\label{19eqa} 4&\leq&
\langle\Psi,\Psi\rangle\lim_{n\to\infty}\left\langle\phi(f_n)\Psi_0,\phi(f_n)\Psi_0\right\rangle\nonumber\\
&=&\langle\Psi,\Psi\rangle\lim_{n\to\infty}\int_0^\infty
\Big[\int_{\R^4} \left|\hat\varphi_n(k^0-\omega)g({\bf
k})\right|^2 \nonumber\\
&\times&\delta^+_s(k)dk\Big]\rho(ds^2)\nonumber\\
\end{eqnarray}
where $\delta_s^\pm(k)=\theta(\pm k^0)\delta(k^2-s^2)$. Here we
have used the K\"allen-Lehmann representation of the two-point
function of $\phi(x)$ with $\rho$ the spectral measure of
$\phi(x)$. Now, the functions $|\hat\varphi_n(k^0-\omega)g({\bf
k})|^2$ converge to zero for $n\to\infty$ and all $k\in\R^4$.
Furthermore, there is a uniform bound for these functions, namely
$|2\chi_{\{-1\leq k^0-\omega\leq 1\}}(k)g({\bf k})|^2$ ($\chi_A$
being the indicator function of the set $A$) and this uniform
bound is integrable w.r.t. the measure
$\int_0^\infty\delta^+_s(k)\rho(ds^2)$. Therefore, the right hand
side of Eq.(\ref{19eqa}) converges to zero by the theorem of
dominated convergence and we have derived a contradiction. This
establishes Theorem 1.

One can summarize the above proof as follows: On shell
singularities $\sim 1/\kappa$ in the first argument of a vacuum
expectation value which lead to decay reactions are continuous
distributions w.r.t. a $C^1$-norm and can not be dominated by  a
measure of K\"allen-Lehmann type which is (locally) continuous
w.r.t. a $C^0$-norm.

It should also be noted that if one drops the assumption Eq.
(\ref{8eqa}) one can still conclude $\langle\phi^{\rm
out}(x)\Psi_0,\Psi\rangle=0\Leftrightarrow\langle\phi^{\rm
in}(x)\Psi_0,\Psi\rangle =0$ and thus ${\cal H}^{\rm in}_1={\cal
H}^{\rm out}_1$ with ${\cal H}_1^{\rm
in/out}=\overline{\{\phi^{\rm in/out}(f)\Psi_0:f\in{\cal S}\}}$.
\section{A QFT with indefinite metric and decaying particles}
Consider the sequence of truncated Wightman functions in
energy-momentum variables given by the standard two point
functions ($\alpha_l=1,2$)
\begin{eqnarray}
\label{Y1eqa}
\left\langle\Psi_0,\hat\phi_{\alpha_1}(k_1)\hat\phi_{\alpha_2}(k_2)\Psi_0\right\rangle^T&=&
\delta_{\alpha_1,\alpha_2}\delta^-_{m_{\alpha_1}}(k_1)\nonumber\\
&\times&\delta(k_1+k_2)
\end{eqnarray}
($\hat \phi_\alpha(k)$ being defined as the Fourier transform of
the quantum fields $\phi_\alpha(x),\alpha=1,2$) and $m_1=\mu$ and
$m_2=m>2\mu$ and three point functions
\begin{eqnarray}
 \label{Y2eqa}
&&\left\langle\Psi_0,\hat\phi_{\alpha_1}(k_1)\hat\phi_{\alpha_2}(k_2)\hat\phi_{\alpha_3}
(k_3)\Psi_0\right\rangle^T \noindent\\ &=&
\prod_{\l=1}^3p_{\alpha_l}(k_l)\sum_{\beta_1,\beta_2,\beta_3=1,2\atop
j=1,2,3}\prod_{l=1}^{j-1} \delta^-_{m_{\beta_l}}(k_l)
\nonumber\\&\times&{1\over k_j^2-m_{\beta_j}^2}
\prod_{l=j+1}^3\delta_{m_{\beta_l}}^+(k_l)
\delta(k_1+k_2+k_3)\nonumber\\
\end{eqnarray}
and all other truncated $n$-point functions equal to zero. Here
$p_1(k)=(k^2-m^2)/(m^2-\mu^2)$, $p_2(k)=(k^2-\mu^2)/(m^2-\mu^2)$.
One can prove that (i) the so-defined Wightman functions fulfill
the requirements of temperedness, Poincar\'e invariance, locality,
Hermiticity, clustering and spectrality from the Wightman
axioms\cite{SW}, see \cite{AGW1,Go1,Go2}; (ii) they are the vacuum
expectation values of a QFT with indefinite metric in the sense of
\cite{MS}, see \cite{AGW2,Go1}; (iii) asymptotic fields for this
theory can be constructed and the LSZ asymptotic condition holds
in the form of Yang-Feldman equations, cf. \cite{AG,Go1}; the
transition amplitude for the reaction $m\to \mu+\mu$ for this
model is given by
\begin{equation}
\label{X3eqa} 2\pi i
\delta_m^+(k_1)\delta_\mu^+(k_2)\delta_\mu^+(k_3)\delta(k_1-(k_2+k_3)).
\end{equation}
 cf. \cite{AG,Go1}. Thus, all requirements of Section 2
and 3 with the exception of positivity are not in contradiction
with the decay reaction (\ref{1eqa}). This shows that positivity
is really crucial for Theorem 1 and that the energy-time
uncertainty relation can be violated in the case of QFTs with
indefinite metric.

\

 \noindent {\bf Acknowledgments.} I would like to thank
  B. Kuckert and G. Morchio for helpful discussions. The referee's
suggestions for the revision of the article and
 financial support by ``Hochschulsonderprogramm III'' of the FRG via a DAAD scholarship
 are gratefully acknowledged.

\end{document}